\documentclass[aps,pr1,twocolumn,groupedaddress]{revtex4-1}
\usepackage{graphicx}
\usepackage{dcolumn}
\usepackage{amsmath}
\begin{document}

\title{Cubic MnSb: epitaxial growth of a predicted room temperature half-metal}

\author{James D. Aldous}
\author{Christopher W. Burrows}
\author{Ana M. S\'anchez}
\author{Richard Beanland}
\author{Ian Maskery}
\author{Matthew K. Bradley}
\author{Manuel dos Santos Dias}
\author{Julie B. Staunton}
\author{Gavin R. Bell}
\email[Electronic mail:]{  gavin.bell@warwick.ac.uk}
\affiliation{Department of Physics, University of Warwick, Coventry, UK, CV4 7AL}

\date{\today}

\begin{abstract}
Epitaxial films including bulk-like cubic and wurtzite polymorphs of MnSb have been grown by molecular beam epitaxy on GaAs via careful control of the Sb$_{4}$/Mn flux ratio. Nonzero-temperature density functional theory was used to predict \textit{ab initio} the spin polarization as a function of reduced magnetization for the half-metals NiMnSb and cubic MnSb. In both cases, half-metallicity is lost at a threshold magnetization reduction, corresponding to a temperature $T^{*} < T_{C}$. This threshold is far higher for cubic MnSb compared to NiMnSb and corresponds to $T^{*} > 350$K, making epitaxial cubic MnSb a promising candidate for room temperature spin injection into semiconductors. 
\end{abstract}

\maketitle

Efficient electrical spin injection is crucial for the development of semiconductor (SC) spintronics, in which spin-polarized charge carriers flow from a ferromagnetic (FM) material into a non-magnetic SC structure. An ideal FM material would provide a fully spin polarized current \cite{HMFreview:2007} and should possess ``engineering compatibility'' with its SC host, which includes controllable epitaxy and interfaces, and suitability for device processing. Certain half-metallic ferromagnetic (HMF) materials have a density-of-states (DOS) spin polarisation at the Fermi level of $P_{DOS}$ = 100\% due to the presence of a band gap straddling the Fermi energy $E_F$ for one spin channel only \cite{HMFspecissue:2007}. Density functional theory (DFT) has predicted this property in several classes of materials \cite{HMFreview:2007}, including Heusler alloys and transition metal pnictides (TMPs), at temperature $T=0$K. On this basis, half-metallicity was first envisaged in NiMnSb, which has a minority spin gap $E_{g} \approx $~0.5 eV \cite{deGroot:1983}. However, there is presently no HMF candidate with SC engineering compatibility which clearly maintains $P_{DOS}\approx$ 100\% at room temperature. 

Experimental verification of $P_{DOS}$ is not straightforward and after more than two decades' study, $P_{DOS}(T)$ for NiMnSb is not known. Positron annihilation experiments on bulk NiMnSb at $T=27$K support $P_{DOS}\approx$ 100\% \cite{positron}. Spin polarized photoemission spectroscopy (SPPES) can, in principle, measure $P_{DOS}$ more directly but it is strongly affected by the magnetic, chemical and structural conditions of the surface, which may differ from those of the bulk. Threshold SPPES measured $P_{DOS}\leq$ 50\% at $T=$ 20K \cite{PES1985}, while more recent works measured 40\% \cite{PES2001} or 50\% at $T=$ 10K falling to 40\% at $T=$ 300K \cite{PES2006}. Inverse SPPES experiments implied similar $P_{DOS}$ values, again more dependent on surface condition than $T$ \cite{SPIPES}. Note that spin polarizations in electron transport are not the same as $P_{DOS}$, and have been inferred from tunnel junctions \cite{tunnel} or spin light emitting diodes \cite{spinLED}, or by Andreev reflection \cite{Andreev2006}. For structures incorporating NiMnSb, polarizations between 2\% and 30\% are typical, even at low temperatures.

Given the difficulties of measuring $P_{DOS}$ unambiguously, theoretical predictions play a key role in assessing the potential of HMF materials. The binary TMPs have attracted much recent attention. This class includes materials such as MnAs and CrSb which are predicted to be wider-gap HMFs in cubic (sphalerite) \cite{cubic:2000, cubic:2003} and wurtzite \cite{wurtzite} (w-) structures. For example, we predict cubic c-MnSb to be HMF with $E_{g} =$ 1.08 eV, while n-MnSb (its naturally occurring niccolite B8$_{1}$ structure) is a normal FM metal. Attempts have been made to stabilize cubic polymorphs by high-strain epitaxial growth on common SC substrates, \textit{e.g.} c-CrAs and c-CrSb grown as ultra-thin films on GaAs(001). However, these films revert to their stable non-HMF structures within 2-4 nm thickness \cite{cubic:2000, CrAs:2001} while X-ray diffraction (XRD) has shown the presence of strained orthorhombic o-CrAs rather than c-CrAs \cite{CrAsXRD}. DFT calculations predicted that biaxial strain would rarely stabilize such HMF polymorphs \cite{biaxial}. Furthermore, a DFT study of CrAs ultra-thin films on GaAs(001) found that bulk-like o-CrAs was favorable for films with just 3 Cr atomic layers, and argued that reported FM hysteresis is due to uncompensated interfacial spins of anti-FM o-CrAs \cite{CrAs:2010}. One concludes that highly strained epitaxy is not a promising route to producing wide-gap HMF thin films of this class.

While extrinsic effects can reduce spin polarization, such problems can be tackled by improving material quality \cite{NiMnSb:2010} and epitaxy \cite{Interfaces:2006}. It is crucial to address the \textit{intrinsic} limitations on $P_{DOS}$. It has been realised that HMF band structures change as $T$ increases, even well below the Curie temperature $T_{C}$ \cite{HMFReview:2008, SpinPolT:2006, SpinPolT:2008, SpinPolT:2004}. The DOS in the minority spin gap becomes nonzero as the thermal fluctuations produce disorder in the local spin moment alignment. In fact, beyond a threshold temperature $T^{*} < T_{C}$, the value of $P_{DOS}$ collapses far more rapidly than the reduction of magnetisation $M(T)$ \cite{SpinPolT:2006}. 

We report here the successful stabilization of bulk-like c-MnSb and w-MnSb crystallites within fully relaxed n-MnSb films grown by molecular beam epitaxy (MBE) on GaAs(111). We confirm by DFT that these are HMF materials at $T=0$K. However, we also calculate $P_{DOS}$ as a function of magnetization $M(T)$ using nonzero-temperature DFT, and show that $P_{DOS}$ in c-MnSb is much more robust against finite temperature effects than in NiMnSb. 

\begin{figure}[!t]
\includegraphics[scale=0.33]{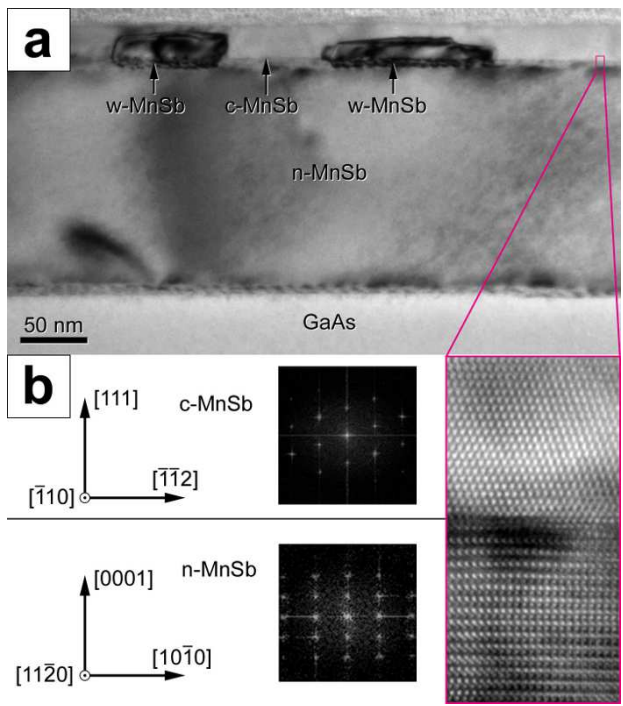}
\caption{\label{TEM} (color online) TEM data for polymorphic MnSb films on GaAs(111). A typical bright-field image (a) and Fourier transforms from individual regions (b) are labelled with constituent phases, while a high resolution micrograph of the c-MnSb/n-MnSb boundary shows a sharp epitaxial interface.}
\end{figure}

MnSb films were grown on $\sim$10~mm wide GaAs(111) substrates by MBE, without substrate rotation \cite{Hatfield:2007, Hatfield:thesis}. The substrate temperature $T_{sub}$ was fixed at $695\pm5$~K, and the growth rate was 6~nm min$^{-1}$. For each growth run the Sb$_{4}$/Mn flux ratio $J$ was measured directly by a retractable ionization gauge. Typically, samples were capped with Sb to prevent oxidation \cite{Hatfield:2009}. Specimens with varying total film thickness and $J$ were prepared for transmission electron microscopy (TEM) using conventional techniques. They were examined using JEOL 2000FX and Tecnai F20 microscopes operating at 200~kV, using bright field and dark field imaging as well as selected area diffraction patterns (SADP). 

In Fig. \ref{TEM} we show TEM data from typical polymorphic films. Initially the epilayer grows as n-MnSb, with a high density of misfit dislocations at the n-MnSb/GaAs interface efficiently relaxing the 3.2\% in-plane strain within a few nm thickness \cite{Hatfield:2007, Hatfield:thesis}. However, beyond a certain film thickness (typically 100 nm) a sharp transition to polymorph growth occurs. Careful analysis of the images and SADP reveals that two epitaxial polymorphs of MnSb exist in the upper portion of the films, namely c-MnSb(111) and w-MnSb(0001). Calibrating SADP data to the GaAs substrate gives the following lattice parameters (in {\AA},  uncertainty $\pm$0.5\%): n-MnSb \textit{a}=4.115, \textit{c}=5.769; w-MnSb \textit{a}=4.291, \textit{c}=7.003; c-MnSb \textit{a}=6.502. For n-MnSb, the lattice parameters conform with results in the literature, which gives us confidence in the w-MnSb and c-MnSb values, while for w-MnSb, the $c/a$ ratio is almost exactly $1.632=\sqrt{8/3}$, corresponding to tetrahedral coordination of nearest neighbours. For the substrate and all three polymorphs, the half-crystal symmetries are 3m and in the absence of translational domains a single epitaxial orientation is predicted \cite{hetero}. We see no TEM evidence for translational domains and associated double positioning boundaries in the epilayer due to pre-existing GaAs surface steps. The polymorph growth is not influenced by post-growth cycling between 300~K and 700~K and the film structure is stable for many months, provided the samples are not heated outside ultra-high vacuum and are protected from oxidation \cite{Hatfield:2009}. Furthermore, XRD data (not shown) from as-grown films show the presence of w-MnSb and c-MnSb in addition to n-MnSb, and so we can rule out polymorphism as an artefact of TEM specimen preparation. The interface between n-MnSb and the polymorph epilayer is sharp, as shown by the high resolution image in Fig. \ref{TEM}. 

On the basis of these observations, we attribute the nucleation of polymorphs to surface processes during MBE growth. In fact, the onset of polymorphism in the films is most strongly correlated with the flux ratio \textit{J}: careful control of both $J$ and $T_{sub}$ is required to produce c- and w-MnSb. For $J<5$, the growth rate is Sb-limited \cite{Hatfield:thesis} while for $J>8$ the film becomes faceted and non-planar. In the range $6.6<J<7.5$, c-MnSb crystallites readily appear in the n-MnSb films. We have not observed polymorphism in films grown at $T_{sub} \le 675$~K. Surface segregation of Sb leading to a double-layered surface reconstruction with local tetrahedral coordination is a potential explanation for the transition, which requires a layer stacking order change from ABAC to AaBbCc or AaBbAa in the [0001] or [111] direction. The mechanism for stabilization of these polymorphs is clearly very different from direct epitaxial strain of the SC substrate previously investigated \cite{CrAs:2010, cubic:2000, CrAs:2001, CrAsXRD, biaxial}. XRD shows a predominance of c-MnSb over w-MnSb in most films and we therefore focus on c-MnSb below.

We calculated the $T=0$ electronic properties of the three polymorphs using two standard DFT packages \cite{Munich, CASTEP}. Full details will be given in a longer paper. Both c-MnSb and w-MnSb are HMF materials with $E_{g}$ 1.08~eV and 1.15~eV respectively and integer total spin moment 4$\mu_B$. The calculated electronic structure of n-MnSb agrees with previous work \cite{Rader:1998} and shows a weak $P_{DOS}$ of 18\%. Trends in optimized DFT lattice parameters between the three polymorphs correspond well with the SADP data. The change of MnSb structure from n- to c- or w- represents true polymorphism since there is a change of nearest-neighbour bonding to tetrahedral. Conversely, the c- and w- phases are mutual polytypes, differing only in the stacking sequence of (111) or (0001) planes. As expected, they are very close in total energy, calculated respectively at 0.829 and 0.833 eV per formula unit above n-MnSb. 

\begin{figure}[!t]
\includegraphics[scale=0.33]{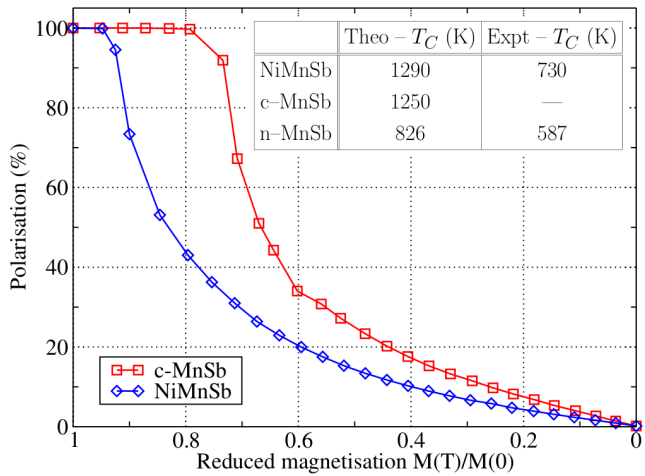}
\caption{\label{SP_redM} (color online) Nonzero-temperature electronic structure theory for MnSb and NiMnSb. The curves show the calculated Fermi level spin polarisation $P_{DOS}$ as a function of magnetization reduction for c-MnSb and NiMnSb, while the inset table summarises calculated $T_{C}$ values from DLM-DFT. The experimental lattice parameter was assumed for c-MnSb, and the value used for NiMnSb matched that of Ref. \cite{SpinPolT:2006}.}
\end{figure}

We have extended the modelling of the c-MnSb and n-MnSb polymorphs to nonzero-temperature DFT using the disordered local moment (DLM) theory \cite{Julie:24}, implemented using the Korringa, Kohn, Rostoker (KKR) multiple-scattering method \cite{Julie:25, Julie:26}, to examine the robustness of the half-metallicity. The theory includes the effects of thermally induced spin fluctuations on the underlying electronic structure, which must eventually destroy the long-range magnetic order and the overall spin polarization of the electronic states above the Curie temperature $T_C$. The DLM-DFT approach goes beyond the usual low temperature approximation of preset spin polarization and associates local spin polarization axes with each lattice site. These are allowed to vary slowly on the timescale of the electronic motions, affecting the nature and extent of the spin polarization of the electronic structure and being self-consistently maintained. Thermally-induced disorder in the orientations of the local moments is treated using a mean-field scheme, based on the coherent potential approximation \cite{Julie:35, Julie:36}. By taking appropriate ensemble averages over the local moments' orientational configurations, we are able to determine the system's magnetic properties and find the temperature dependence of the spin polarized electronic structure and hence $P_{DOS}$. This DLM-DFT theory also describes \textit{ab initio} the onset and type of magnetic order \cite{Julie:24, Julie:25, Julie:26, Julie:27, Julie:28, Julie:29}. The expression for the paramagnetic spin susceptibility, $\chi(\mathbf{q},T)=\frac{\mu^2 }{3k_{B}T-S^{(2)}(\mathbf{q},T)}$, from which $T_{C}$ values are estimated, is derived by considering the response of the DLM paramagnetic state to the application of an external, site-dependent, magnetic field.  Formally, $S^{(2)}(\mathbf{q},T)$ is the direct correlation function for the local moments, of size $\mu$, and is the analogue of an exchange integral in the classical Heisenberg model of interacting spins. An important difference, however, is that $S^{(2)}(\mathbf{q},T)$ contains a fully itinerant description of electrons, whereas in the Heisenberg model the exchange integral is defined in terms of interactions between localized spins. Details of the methods used to evaluate $S^{(2)}$ can be found in refs. \cite{Julie:28, Julie:29}.

Results of our DLM-DFT calculations are summarised in Fig. \ref{SP_redM}. The inset tabulates the Curie temperatures for n-MnSb and NiMnSb: our calculated mean-field $T_{C}$ values are larger than the experimental results but with the correct comparative magnitudes between the two materials. Previous calculations \cite{SpinPolT:2006} for NiMnSb have also somewhat overestimated $T_{C}$. Our calculated values for NiMnSb and c-MnSb are very similar. That $T_{C}\gg$~300~K for c-MnSb is promising but, as pointed out previously, does not guarantee a high $P_{DOS}$ at room temperature \cite{SpinPolT:2004}. The behavior of $P_{DOS}$ predicted by DLM-DFT as a function of magnetization reduction is plotted in Fig. \ref{SP_redM} for c-MnSb and compared with NiMnSb: the horizontal axis is effectively a distorted temperature scale between 0 and $T_{C}$. For NiMnSb, $P_{DOS}$ rapidly collapses even after around 5\% reduction in magnetization. The same behavior is evident for c-MnSb but in this case $P_{DOS}$ remains at 100\% to a much higher magnetization reduction of around 21\%, essentially due to the much larger minority spin gap. Using DLM-DFT to estimate the magnetization dependence on temperature $M(T)$ we can compare $T^{*}$ of NiMnSb and c-MnSb \cite{Julie:27}. We find that while for NiMnSb $T^{*} \geq$ 100~K, in c-MnSb $T^{*} \geq$ 350~K, comfortably above room temperature. These values are lower-bound estimates since the theory typically overestimates the rate of decrease of $M$ with rise in $T$; they are based on $T_{C} \approx$ 730~K rather than the higher calculated values.

In c-MnSb the nearest neighbours of each Mn atom are 4 Sb atoms. The local tetrahedral environment enables the $t_{2g}$ states of Mn to hybridise easily with the Sb p-states so that a large bonding-antibonding splitting is set up where the lower-lying bonding states have more p-character whilst the higher antibonding states are more d-like. The gap in between is partly filled by the non-bonding narrow $e_{g}$ states of the transition metal. Mn has a large exchange splitting which causes the bands of the majority and minority spin electrons to be distributed around different energies \cite{Galanakis:2003}.  At $T=0$ the majority spin d-states of the Mn atoms are closer to the Sb p-states and the bonding-antibonding gap is smaller. For the minority spin electrons the Mn d-states are forced higher by the exchange splitting and the gap for this spin channel straddles $E_F$ with the anti-bonding states unfilled. 

\begin{figure}[!t]
\includegraphics[scale=0.2]{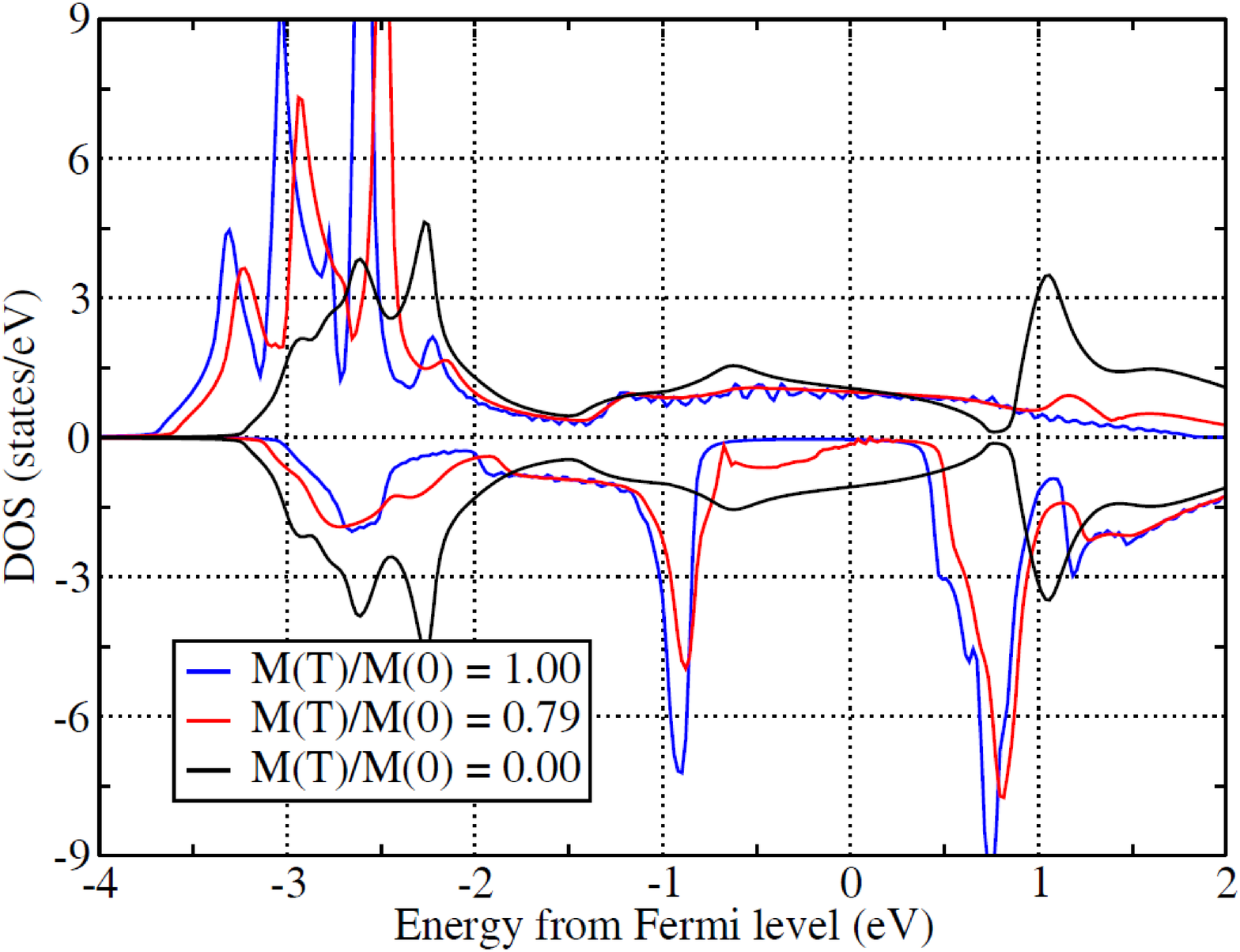}
\caption{\label{SPDOS} (color online) Spin-polarized DOS for c-MnSb at different magnetizations $M(T)$ calculated by DLM-DFT at the experimental lattice parameter. The three curves shown correspond to $T=0$K (blue), $T=T^{*}$ (red) and $T=T_C$ (black).}
\label{SPDOS}
\end{figure}

As the temperature is raised a local exchange splitting of the electronic structure persists even above $T_{C}$ despite the orientations of the local spin polarization directions becoming disordered. A simple weighting of the majority and minority spin electronic structure to reflect the decrease in overall spin polarisation would produce a linear dependence of  $P_{DOS}$  on $M(T)$. This is not seen in Fig. \ref{SP_redM}, however, and to quantify $T^{*}$ we have to assess the effects of the local moment disorder on the electronic structure more carefully. This is illustrated for c-MnSb in Fig. \ref{SPDOS}. The minority spin gap $E_{g} =$ 1.08 eV appears in the fully magnetized case corresponding to $T=0$K (blue curve). The separation between the antibonding $t_{2g}$ states of the majority and minority spin channels is large so that as the overall magnetization is reduced, ``impurity'' states associated with minority spin electrons appear in the majority spin channel above $E_F$ whilst majority spin impurity states form in the minority spin gap close to the position of the antibonding $t_{2g}$ majority spin band. As $M$ is reduced, these encroach up to the Fermi level and at $T^{*}$ remove the half-metallicity abruptly. This is shown in the red curve of Fig. \ref{SPDOS}: the impurity states have just reached $E_F$ at a magnetization reduction of 21\%, initiating the collapse of $P_{DOS}$ shown in Fig. \ref{SP_redM}. Finally, for $T=T_C$ all spin polarization is lost (black curve). For NiMnSb, the same effects occur but $T^{*}$ corresponds to a magnetization reduction of just 5\% due mainly to the smaller value of $E_g$.

The mechanism of spin polarization reduction in hybridization-gap materials has been confirmed by the present fully \textit{ab initio} calculations and elsewhere \cite{HMFReview:2008, SpinPolT:2006, SpinPolT:2008, SpinPolT:2004}. It is clear that, as well as $T_C$, $E_g$ is a key parameter in determining the high-temperature spin polarization. Gaining full control of the MBE growth of TMP polymorphs with large gap will be an essential step towards exploiting these new materials. Fully \textit{in situ} investigations of the transition to polymorphic growth should shed more light on the microscopic surface processes leading to the growth transition. The combination of engineering compatibility, large minority spin gap and robust spin polarisation make c-MnSb highly attractive for spintronics applications. Epitaxial interfacing to a semiconductor remains a crucial task but the prognosis here is also good. Mollet and Jenkins \cite{Jenkins:2007} have calculated the (001) and (111) surface electronic structure of c-MnSb using DFT and found that half-metallicity is preserved for the favoured Sb-terminated surfaces. Since this structure is shared with common semiconductors such as GaAs, it is likely that half-metallic interfaces can readily be formed. In contrast, the preservation of half metallicity at SC interfaces is much less common with ternary Heusler alloys \cite{Interfaces:2006}. Interestingly, Jenkins's recent prediction of tilted Dirac cones (similar to those found in graphene and at topological insulator surfaces) in the surface states of NiMnSb \cite{Jenkins:2011} is also likely to translate to c-MnSb, since the relevant symmetry of the structures is identical. 

We acknowledge support from EPSRC, the Royal Society, Birmingham Science City: Creating and Characterising Next Generation Advanced Materials, and Advantage West Midlands and the European Regional Development Fund, plus a PhD grant SFRH/BD/35738/2007 awarded by FCT Portugal using funding from FSE/POPH.

\end{document}